\documentclass[12pt]{article}

\def\be{\begin {equation}}

\def\ee{\end {equation}}

\def\bea{\begin{eqnarray}}

\def\eea{\end{eqnarray}}

\begin{document}

\vspace{5cm}

\thispagestyle{empty}

\begin{center}

{\LARGE \bf The thermodynamic limit of the Whitham equations}

\vspace{1.0cm}

{\Large G.A. El}

School of Mathematical and Information Sciences, Coventry University, Coventry, UK

\end {center}
\begin{abstract}
The infinite-genus limit of the KdV-Whitham equations is derived. The limit involves special scaling for the associated spectral surface
such that the integrated density of states remains finite as $N \to \infty$ (the thermodynamic type limit).
The limiting integro-differential system describes slow evolution of the density of states and can be regarded as the kinetic equation for soliton gas. 

\end{abstract}

\section{Introduction}
As is well known \cite{Wh}, \cite{FFM},  the modulation (Whitham) equations for integrable systems describe gradual variations 
of spectral Riemann surfaces, whose local structure define $x,t$- oscillating finite-gap potentials. The wave dynamics problems leading to the 
modulation equations are
typically connected with the study of the long-time evolution of  large-scale initial data \cite{LL}, \cite{V87} and  involve
the semiclassical asymptotics of the inverse scattering transform. In this asymptotics, the typical scale of modulations $1/\epsilon$
is necessarily proportional to a global number of degrees of freedom in the problem $g \gg1$ (we assume 
no small parameters in the original equation, instead, we consider large-scale $\sim 1/\epsilon $ initial data that can be approximated by $g$-
soliton or $g$-gap potential). 
Then, asymptotically, the solution manifests itself  as a modulated
$N$-gap potential, where $N \ll g$ \cite{V90}, \cite{DVZ} . The whole point of the traditional modulation theory, thus, 
is to reduce a complicated system with  many degrees of freedom to a simpler one with a few degrees of freedom.

The modulation equations, however, can be considered on their own outside any relation to the semiclassical limit in the 
initial value problem.  Then, at least formally, the question of possibility of the infinite-genus limit for the modulation equations 
can be risen.   In this paper, I show that this limit, along with nontrivial mathematical meaning, has  natural physical interpretation.

The paper is dedicated to the study of   the limit as $N \to \infty$ for the $N$-gap averaged Korteweg -- de Vries (KdV) equation.  
The KdV equation is 
taken in the form
\be \label{KdV}
u_t -6uu_{x} +  u_{xxx} =0 \, . 
\ee
The key in the limiting transition for the modulation equations  is the scaling of the spectral surface
introduced by Venakides 
in the study of the  continuum limit of theta-functions \cite{V89}. The scaling implies that the essential part of the spectrum of the Shr\"odinger operator 
associated with (\ref{KdV}) lies in a finite interval , say,  $(-1,0)$ and the genus of the problem  $N \gg 1$.  
Now, with the width of gaps being  $O(1/N)$ and the bands being exponentially narrow,  $O(\exp (-N))$, 
the limit as $N \to \infty$ for the KdV-Whitham equations  is shown to exist (Section 5).  
The existence of the limit for the Whitham equations is ultimately connected with the existence of the limit for the quantity called a 'rotation number' 
(or 'integrated density of states') introduced for almost periodic potentials by Johnson and Moser in  \cite{JM}. 
Since the established limit  preserves finiteness of the density of states as $N \to \infty$,  it can be called  the thermodynamic limit.
The limit of this type has been considered recently in \cite{EKMV} in connection with stochastic description of infinite-gap potentials.

The main result of the current paper is presented  in Section 6 where the thermodynamic limit for the KdV-Whitham equations is derived in the 
form of a nonlinear transport equation (\ref{kinsol}) for the density of states. The connection between 'density' and 'velocity' in (\ref{kinsol})
is given by the integral equation (\ref{ints}). I suggest that the obtained limit describes macroscopic dynamics of the 'thermodynamic limit of 
stochastic soliton lattices' considered in \cite{EKMV} and can be regarded as the kinetic equation for solitons. This interpretation is also
supported by the exact agreement of the small density asymptotics of the resulting system with Zakharov's  kinetic equation for a rarefied 
soliton gas \cite{Zakh}.  

\section{The Whitham equations: basic definitions}
In this section, we give a brief account on some basic definitions and notations from the finite-gap theory that will 
be used in this paper.
The KdV-Whitham system of genus $N$ can be represented as  one generating equation in the Flashka-Forest-McLaughlin form 
\cite{FFM}
\be \label{Wh}
\partial_T dp_N=\partial_X dq_N \, ,  
\ee
$$ X=\epsilon x \,  , \qquad T=\epsilon t \,  , \qquad \epsilon \ll 1 \,  ,$$
where $dp_N$ and $dq_N$ are the meromorphic differentials on the hyperelliptic Riemann surface of genus $N$
\be \label{rs}
\Gamma : \qquad R ^2 (E)= \prod \limits _{j=1}^{2N+1}(E - E_j) \, , \qquad E \in {\bf C} 
\ee
$$
E_1<E_2< \dots < E_{2N}< E_{2N+1} \, , \qquad E_j \in {\bf R}
$$
with cuts along bands  $[E_{2j-1}, E_{2j}]$.  The canonical basis of cycles  is organized as follows:
the  $\alpha_j$-cycle surrounds the $j$-th cut clockwise on the upper sheet, and the $\beta_j$-
cycle is canonically conjugated to $\alpha_j$'s such that the closed contour $\beta_j$
starts at $E_{2j}$ , goes to $+\infty$ on the upper sheet and returns to $E_{2j}$
on the lower sheet .

Then the meromorphic differentials are defiened by:
\be \label{pn} \
dp_N(E)=\frac{E^N+b_{N-1}E^{N-1}+\dots+b_0}
{R(E)}dE\, , \qquad \oint \limits _{\beta_j}dp_N(E)=0 \, , \qquad j=1,\dots , N \,
\ee

\be \label{qn} \
dq_N(E)=12 \frac{E^{N+1}+c_{N}E^{N}+\dots+c_0}
{R(E)}dE\, ,\qquad c_N=-\frac{1}{2}\sum \limits _{j=1}^{2N+1}E_j
\ee
$$
\oint \limits _{\beta_j}dq_N(E)=0 \, , \qquad j=1,\dots , N \, .
$$
The differentials $dp_N$ and $dq_N$ are often referred to as quasimomentum and quasienergy differentials.
The integrals of $dp_N$ and $dq_N$ over the
$\alpha$ - cycles are known  to give the
components of the wave number and the frequency vectors \cite{FFM}, \cite{DN}
\be \label{k}
 \oint  \limits _{\alpha_j} dp_N(E)=k_j \,  , \qquad
\oint  \limits _{\alpha_j} dq_N(E)=\omega_j \, , \qquad  j=1, \dots , N \
\ee
The finite-gap potentials associated with the Riemann surface (\ref{rs}) are known to be the multiphase solutions of 
the KdV equation (\ref{KdV})  \cite{N}, \cite{L},  \cite{FFM}
\be \label{uy}
u_N(x,t;X,T)=u_N(y_1,  \dots, y_N)\,  , 
\ee
where the real phases 
\be \label{phase}
y_j=k_jx+\omega_j t + f_j\,  , \qquad  j=1, \dots , N \, .
\ee
are  the linear function of 'fast' $x$ and $t$  while the  dependence on 'slow' variables $X$ and $T$, 
enters the potential $u_N$ via the branch points $E_j(X,T)$ of the Riemann surface. 
The constant $f_j(mod 2\pi)$ is the fixed initial phase. 
The function $u_N({\bf y})$ is $2 \pi$-periodic in each phase $y_j$
\be \label {uq}
u_N(y_1, \dots, y_j+2\pi, \dots, y_N )=
u_N(y_1, \dots, y_j, \dots, y_N )\, ,
\ee
that is $u_N(x,t)$ for  fixed $E_1, \dots E_{2N+1}$ is $N$-quasiperiodic both in $x$ and $t$ 
(hereafter we consider only incommensurable $k_j$s and $ \omega_j $s).

The Whitham equations (\ref{Wh}) thus describe slow evolution (modulations) of  rapidly oscillating finite-gap potentials.

The algebro-geometric structure of the finite-gap potentials is given by the basis holomorphic differentials
\be \label{psi}
\psi_j = \sum \limits_{k=0}^{N-1}a_{jk} \frac{E^{k}}{ R (E)}dE \, , \qquad j=1, \dots, N .
\ee
and the functions $a_{jk}(E_1, \dots , E_{2N+1})$ are determined by the normalization over the $\alpha$-cycles
\be \label{norm}
\oint \limits_{\alpha _k}\psi _{j} = \delta _{jk}\, ,
\ee
while the integrals over the $\beta$-cycles give the entries of the Riemann matrix $B_{ij}$
\be \label{B}
B_{ij}= \oint \limits _{\beta_j}\psi_i \, . \
\ee
The wavenumbers $k_j$ and the frequences $\omega_j$ are expressed in terms of $a_{ij}$ by the formulas
\cite{IM}, \cite{FFM}
\be \label{kom}
{\bf k} = -4\pi iB^{-1}{\bf a}_{N-1}\, ,
\ee
$$
{\bf \omega} = (\omega_1, \dots,
\omega_N ) = -8\pi i B^{-1} ({\bf a}_{N-1}\sum \limits_{j=1}^{2N+1}E_j
+ 2{\bf a}_{N-2}) \, .
$$
\section{Rotation numbers and the density of states}
We consider the Schr\"odinger equation  
\be \label{schr}
(-\partial ^2 _{xx}+ q(x))\psi=E \psi \, ,
\qquad  x \in {\bf R} \, .
\ee
with an almost periodic  potential $q(x)$. 
The potential $q(x)$ has an important characteristics,
{\it the rotation number}, which was defined 
 by  Johnson and Moser  \cite{JM} as
\be \label{rot}
 \alpha(E)=\lim \limits _{x \to \infty} \frac{1}{x}
 \arg (\psi' (x, E) + i\psi(x,E))\, , \qquad E \in {\bf R}
 \ee
As shown in \cite{JM}, this agrees with the value $\pi \cal N$ where ${\cal N}(E)$ is the
{\it  integrated density of states} defined by  
%We will use also the {\it integrated density of states} ${\cal N}(E)$
%which is connected with the rotation number by a 
%simple relation: 
%The integrated density of states can be also defined directly (Johnson, Moser 19...)
\be \label{ds}
{\cal N}(E)=\lim \limits_{|b-a| \to \infty}\frac{\nu(a,b;E)}{b-a} \, ,
\ee
where $\nu(a,b;E)$ is the number of eigenvalues $E_j \le E$ in the Dirichlet problem on $a\le x \le b$: $\psi(a, E)=\psi(b, E)=0$.
Thus,
\be \label{dens}
{\cal N}(E)= \frac{1}{\pi} \alpha(E) \, .
\ee
If, as a particular case, $q(x)$ is $N$-gap potential , $q(x)=u_N(x)$, such that $-1\le u_N \le 0$
then the rotation number coincides with the real part of quasimomentum \cite{JM}:
\be \label{ap}
\alpha_N(E)=Re \int \limits^{E}_{-1} {dp_N(E')}\, , \qquad E \in (-1,0)\, ,
\ee
Then, with the aid of (\ref{k}) we have
\bea
&& \frac{1}{2}\sum _{j=1}^{M(E)} k_j \qquad \qquad \qquad
\qquad  \  \  \hbox{if} \ \ E \in
\hbox{gap}_M \nonumber \\
\alpha_N (E)&=& \label{alpha}  \\
&& \frac{1}{2}\sum _{j=1}^{M(E)} k_j +
\int \limits^{E}_{E_{2M-1}} dp_N(E') \qquad \hbox{if} \ \ E \in
\hbox{band}_M \, ,\nonumber
\eea
where $M(E)$ is the number of the band nearest to $E$
from the left; $M(E) \le N\, , \ M(0)=N\, , \ M(-1)=0$. 

We note that the formula (\ref{alpha}) agrees with the general statement
in \cite{JM} that the values of the function $2\alpha(E)$ if $E \in
\{\hbox{gap}\}$ belong to the frequency-module of the almost periodic
potential. One can also see that the function $\alpha'(E)$ is positive and
is supported on the spectrum of the finite-gap potential.

Since the finite-gap potential (\ref{uy}), (\ref{phase}) is an almost periodic
function  in  $t$ as well as in $x$
one can also formally introduce a temporal analog of the rotation number
\be \label{beta}
\beta_N  (E)=  Re \int \limits^{E}_{-1} {dq_N(E')}\, ,
\ee
where $dq_N(E)$ is the quasienergy differential (\ref{qn}).

We also introduce the {\it full integrated density  of states} 
$\kappa$ which is a number:
\be \label{kappa}
\kappa =  {\cal N}_N(0)=
\frac{1}{2\pi}\sum _{j=1}^N k_j \, .
\ee
One can see that the full integrated density of states has for the finite-gap potentials
the natural significance of the mean number of waves per unit length.

Analogously, we introduce
\be \label{Omega}
\Omega=\frac{1}{\pi}\beta_N(0)=\frac{1}{2\pi}\sum \limits_{j=1}^{N} \omega_j \, .
\ee
\section{Continuum limit of the spectrum and its thermodynamic nature}

Let the nontrivial (finite band) part of the spectrum lies in the interval
$(-1, 0)$ of the real $E$-axis: $\alpha_N(-1)=0\, ,\ \alpha_N(0)=\pi \kappa$.
For $N\gg1$ we consider the special  band/gap distribution (scaling): 
\be
\label{scal} \hbox{gaps}(E) \sim \frac{1}{N} \, \qquad \hbox{bands}(E) \sim
\exp{(-N)}\, \qquad N\gg1 \, , \ \  E \in (-1,0)\, .
\ee
Following Venakides \cite{V89} we introduce the lattice of points
\be \label{lat}
1\approx \eta _1>\eta _2>\ldots >\eta _N\approx 0\, ,
\ee
where
$$
-\eta _j^2=\frac 12\left( E_{2j-1}+E_{2j}\right)\,
$$
are centers of bands.

We define two positive continuous functions on $[0,1]$:

1. The normalized density of bands $\varphi(\eta)$: 

$$\varphi(\eta)d\eta \approx
\frac{\hbox{number of lattice points in} \  (\eta,  \  \eta + d\eta)}{N}\, .$$ 
That is,
\be \label{phi}
\varphi(\eta_j)=
\frac{1}{N(\eta_j - \eta_{j+1})} +O(\frac{1}{N})\, ,
\qquad \int \limits_0^1
\varphi(\eta)d\eta =1\, , \ \ \eta^2 = -E \in (0, 1) \, .
\ee

2. The normalized logarithmic band width $\gamma(\eta)$:
\be\label{gam}
\gamma(\eta_j)= -\frac{1}{N} \log \delta_j+ O(\frac{1}{N})\,, \qquad \delta_j=E_{2j}-E_{2j-1}\, .
\ee
The functions $\varphi (\eta)$ and $\gamma(\eta)$  asymptotically define the
{\it local} structure of the Riemann surface $\Gamma$ (\ref{rs}) for $N \gg 1$.  In other words, instead of
$2N+1$ discrete parameters $E_j$ we have two continuous functions of $\eta$ on $(-1,0)$, which do not depend
on $x,t$ on $\Delta x \sim \Delta t \sim 1$. 
On a larger scale,
$\Delta x \sim  \Delta t\sim 1/\epsilon$ , $\epsilon \ll 1 $,  in the spirit of the modulation theory, we put 
\be \label{}
\varphi = \varphi(\eta; X,T)\, \qquad \gamma=\gamma(\eta;X,T)\, ,
\ee
where $X=\epsilon x$, $T=\epsilon t$. 
These functions are now subject to the evolution equations which naturally
should be derived as the continuum limit of the Whitham equations (\ref{Wh}) considered for the
same exponential spectral scaling (\ref{scal}).

In the next section  we will show that the scaling (\ref{scal}) implies
the following asymptotic behaviour for the wavenumbers and the frequences 
(these asymptotics can also be derived directly from (\ref{k})): 
\be \label{as}
k_j \sim \omega_j \sim \left| \frac{1}{\log \delta_j}\right |=O(\frac{1}{N}) \, ,  \qquad N \gg 1 \, ,
\ee
\be
\int \limits^{E}_{E_{2j-1}}d p_N(E') \sim \int \limits^{E}_{E_{2j-1}}d q_N(E')
<O( \frac{1}{N})\, , \qquad \,  E \in (E_{2j-1}, E_{2j})\, ,
\ee
which provides  finiteness of the rotation number $\alpha_N(E)$
(and of the integrated density of states ${\cal N}_N(E)$) as
$N \to \infty$:
\be \label{aturb}
\alpha(E) \equiv
\lim \limits_{N \to \infty} \alpha_N(E)=
\lim \limits_{N \to \infty}\frac{1}{2}\sum _{j=1}^{M(E)} k_j < \infty
\, ,\qquad  M \le N \,.
\ee
Due to this property we call the scaling (\ref{scal}) and the
corresponding limit as $N \to \infty$ the {\it thermodynamic} ones. This type of limits
in the finite-gap theory has been introduced in \cite{EKMV}.
Hereafter we will denote the limit of some functional $f[u_N]$ as $N \to \infty$ 
on the thermodynamic spectral scaling as
$$
T\hbox{-}\lim f[u_N]
$$
It should be noted that the thermodynamic spectral scaling (\ref{scal})
appears when one considers the semiclassical asymptotics of the
spectrum for periodic potentials \cite{WK}, \cite{V87},
\cite{V89}. However, in the works just cited, $N \sim 1/\epsilon$, which is the standard
relation in the asymptotics of this type (see also \cite{LL}).  Contrastingly, in the thermodynamic limiting transition,  
the large parameter $N$ defining the scaling  is contained only in the band-gap
structure and is not connected with the parameter $1/\epsilon$, which scales
modulations of the finite-gap potential $u_N(x,t)$.

\section{The thermodynamic limit for the rotation numbers: basic integral equations }
Our aim now is to deduce the thermodynamic limit for the modulation equation (\ref{Wh}). For that, we need to
evaluate the thermodynamic limits of the meromorphic differentials $dp_N$ and $dq_N$. First we make an 
estimate for their imaginary parts on the thermodynamic scaling (\ref{scal}):

\bea
&& \  \ 0 \qquad \qquad \qquad
\qquad  \  \   \ \ E \in
\hbox{any band} \nonumber \\
\left| Im \int \limits^{E}_{-1} dp_N(E')\right|&=& \label{im} \\
&& 
\left |\int \limits^{E}_{E_{2j}} dp_N(E')\right| <O( \frac{1}{N}) \qquad \ \ E \in
\hbox{gap}_j \, .\nonumber
\eea
Therefore,  the imaginary part of $dp_N(E)$ vanishes as $N \to \infty$ and we have for $E \in (-1,0)$:
\be \label{tp}
T\hbox{-}\lim p_N (E)=\alpha(E)=T\hbox{-}\lim \frac{1}{2}\sum \limits^{M(E)}_{j=1} k_j \, .
\ee
Analogously, we get
\be \label{tq}
T\hbox{-}\lim q_N (E)= \beta(E)=T\hbox{-}\lim \frac{1}{2}\sum \limits^{M(E)}_{j=1} \omega_j \, ,
\ee
where $\beta(E)=T\hbox{-}\lim \beta_N (E)$.

Now we evaluate the wave numbers $k_j$ and the frequencies $\omega_j$ for the chosen spectral scaling.
For that, we will make use of the following Venakides observation \cite{V89}: 
for the  spectral distribution
(\ref{scal}) the following approximation for the basis holomorphic
differentials is valid:
\be \label{appsi}
\psi_j \approx -\frac{\eta_j}{2\pi}\frac{\prod \limits^{N}_{i=1, i\ne j}(E+\eta_j^2)}
{\left(\prod \limits^{2N+1}_{k=1}(E-E_j)\right)^{1/2}}dE .
\ee
One can see that
$$
(E-E_{2j-1})(E-E_{2j})=(E+\eta_j^2) + O(e^{-N}).
$$
Then an extensive cancellation becomes possible in (\ref{appsi}) so that the hyperelliptic integrals
in (\ref {norm}), (\ref{B}) turn into elementary ones. As a result, we have the following asymptotics (see \cite{V89} for details):
\be \label{Bij}
B_{ij} \approx -\frac{i}{\pi}\left(\log \left| \frac{\eta_i-\eta_j}{\eta_i+\eta_j}\right| +N\gamma(\eta_i)\delta_{ij}\right)\, , 
\qquad i,j = 1, \dots, N\, .
\ee
\be \label{ca}
{ a}_{j,N-1} \approx -\frac{\eta_j}{2\pi } \, , \qquad
 a_{j, N-1}\sum \limits_{j=1}^{2N+1}E_j + 2 a_{j,N-2} \approx
\frac{\eta_j^3}{\pi }\, .
\ee
Now we represent the general relationships (\ref{kom}) between the real and imaginary periods
of theta function in the form convenient for the thermodynamic limiting transition
\be \label{kB}
B{\bf k}= -4\pi i {\bf a}_{N-1}
\ee
\be \label{oB}
B{\bf \omega} = -8\pi i  ({\bf a}_{N-1}\sum \limits_{j=1}^{2N+1}E_j
+ 2{\bf a}_{N-2}) \, .
\ee
It follows from the asymptotics (\ref{Bij}), (\ref{ca}) that to provide a balance of terms in (\ref{kB}), (\ref{oB})
the following scaling is necessary  : 
$k_j \sim \omega_j \sim O(1/N)$. 
Therefore, we introduce two continuous on $(0,1)$ functions $k(\eta)$ and $\omega(\eta)$ by
\be \label{conkom}
k_j=\frac{1}{N}k(\eta_j)\, , \qquad \omega_j=\frac{1}{N}\omega(\eta_j)\, .
\ee
Then, substituting (\ref{Bij}), (\ref{ca}), (\ref{conkom}),  into (\ref{kB}), (\ref{oB}) we get 
two algebraic systems for $k(\eta_i)$, $\omega(\eta_i)$,  $i=1, \dots , N$
\be \label{sum1}
\sum \limits^{N}_{j=1}\frac{1}{N}\log \left| \frac{\eta_i-\eta_j}{\eta_i+\eta_j}\right| k(\eta_j)
+\gamma(\eta_i)k(\eta_i)=-2\pi \eta_i \, .
\ee
\be \label{sum2}
\sum \limits^{N}_{j=1}\frac{1}{N}\log \left| \frac{\eta_i-\eta_j}{\eta_i+\eta_j}\right| \omega(\eta_j)
+\gamma(\eta_i)\omega(\eta_i)=8\eta_i^3\, .
\ee
Using (\ref{conkom}) we represent the relationships (\ref{tp}), (\ref{tq}) for the
thermodynamic limit of the rotation numbers in a continuum form:
\be \label{conta}
2d\alpha(-\eta^2) = \varphi(\eta)k(\eta) d\eta \, , \qquad 2 d\beta(-\eta^2) = \varphi(\eta)\omega(\eta) d\eta \, ,
\ee
Now one readily  obtains the thermodynamic limit of (\ref{sum1}), (\ref{sum2}) :
\be \label{inta}
\int\limits_0^1 \log \left|\frac{\eta - \mu}
{\eta+\mu}\right|\frac{\mu}{\eta}\alpha'(-\mu^2)d\mu  + \sigma(\eta)\alpha'(-\eta^2)
=\frac{\pi}{2}\, .
\ee
\be \label{intat}
\int\limits_0^1 \log \left|\frac{\eta - \mu}
{\eta+\mu}\right|\frac{\mu}{\eta}\beta'(-\mu^2)d\mu  + \sigma(\eta)\beta'(-\eta^2) 
=-2\pi\eta^2 \, ,
\ee
where 
\be \label{sigma}
\sigma(\eta)=\frac{\gamma(\eta)}{\varphi(\eta)}\, .
\ee
We see that the thermodynamic limit  of the rotation numbers 
is determined by the only function $\sigma(\eta)$ defined on $(0,1)$  (instead of the original two:
$\varphi(\eta)$ and $\gamma(\eta)$). 

The integral equations
(\ref{inta}), (\ref{intat}) have been derived in \cite{EKMV} in connection with the establishing the thermodynamic limit
for the stochastic processes generated by the finite-gap potentials (stochastic soliton lattices).
We note that these  integral equations also appear in  \cite{GMZ} where they determine the 
Lax-Levermore type minimizer for the $N$-soliton solution with randomly distributed soliton phases as $N \to \infty$. 
It is clear that there should be a direct  connection between  the 'stochastic' version of the  Lax-Levermore variational 
problem and the thermodynamic limit of the rotation number. 

\section{Thermodynamic limit of the Whitham equations}
The modulation equation (\ref{Wh}) describe slow $X,T$- deformations of the spectral surface (\ref{rs}).
Now we are going to apply the
thermodynamic limiting transition to (\ref{Wh}). As follows from (\ref{tp}), (\ref{tq}), the local thermodynamic limits are 
\be \label{tlim}
T\hbox{-}\lim dp_N (E)=d\alpha(E)\, \qquad
T\hbox{-}\lim dq_N (E)=d \beta(E) \, .
\ee
We put that on a larger scale
\be \label{}
d\alpha=d\alpha(-\eta^2; X,T)\, \qquad d\beta = d\beta(-\eta^2;X,T) \, .
\ee 
Then the thermodynamic limit of the Whitham system (\ref{Wh}) takes the form:
\be \label{tWh}
\partial_T\alpha'=\partial_X\beta' \, ,
\ee
where the functional dependence $\beta'[\alpha'(-\eta^2; X,T)]$ is given by (\ref{inta}), (\ref{intat}).
The thermodynamic limit of the Whitham equations, therefore, describes evolution of the density of states
$\alpha'(E)/\pi$.

Since $d \alpha (E)>0$ one can introduce a distribution function $f(\eta; X,T)$ by 
\be \label{fras}
f(\eta; X,T)d\eta=\frac{1}{\pi}d\alpha(-\eta^2; X,T) \, , \qquad  \int \limits^1_0 f d \eta = \kappa \, , 
\ee
Introducing
\be \label{s}
s(\eta; X,T)= \frac{\beta'(-\eta^2; X,T)}{\alpha'(-\eta^2; X,T)} \, 
\ee
one can represent (\ref{tWh}) in the form of a nonlinear transport equation:
\be \label{kinsol}
\partial_T f=\partial_X (sf)\, ,
\ee
where the integral equation connecting $s$ with $f$ is obtained   by excluding $\sigma(\eta)$ from (\ref{inta}),  (\ref{intat}):
\be \label{ints}
s(\eta)=-4\eta^2+\frac{1}{\eta}\int \limits^1_0 \log \left|\frac{\eta - \mu}{\eta+\mu}\right|f(\mu)[s(\mu)-s(\eta)]d\mu \, .
\ee
Here, of course, $s(\eta)\equiv s(\eta;X,T)$, $f(\eta)\equiv f(\eta;X,T)$. 
In the case of  small integrated density of states $\kappa \ll 1$ we get 
\be \label{szakh}
s \approx-4\eta^2-\frac{4}{\eta}\int \limits^1_0 \log \left|\frac{\eta - \mu}{\eta+\mu}\right|f(\mu)[\mu^2-\eta^2]d\mu \, ,
\ee
which is Zakharov's expression for the velocity of the `trial' soliton moving through rarefied soliton gas \cite{Zakh}. The integro-differential 
system (\ref{kinsol}), (\ref{ints}), therefore, can be regarded as the finite-density generalisation of the Zakharov kinetic equation for solitons. 

Now we discuss briefly what happens to the underlying finite-gap oscillating microstructure under the thermodynamic limiting transition. 
It has been shown
in  \cite{EKMV} that the thermodynamic limit of the finite-gap potentials should be treated as a {\it random process}. More accurately, 
the thermodynamic limit is defined not for the individual potential  $u_N(x)$ but for the stationary random process generated by considering the 
associated torus equipped with the Haar measure. Such processes were called in  \cite{KE} {\it stochastic soliton lattices}.
The stochastic soliton lattice $\nu(x|\phi)$ is defined by
\be \label{nu}
\nu_N(x|\phi)=u_N(\dots, k_jx+\phi_j, \dots), \qquad j=1,\dots, N \, ,
\ee
where $\phi =(\phi_1, \dots, \phi_N)$ is a random value uniformly distributed on Tor$^N$. 
It was shown in \cite{EKMV} that the thermodynamic limit of the stochastic soliton lattice
\be \label{turb}
T\hbox{-}\lim \nu_N(x|\phi)=\zeta(x-x_j) \, 
\ee
represents a random process defined on the infinite discrete set of points  $\{x_j; j \in {\bf Z}\}$ distributed by Poisson on ${\bf R}$. 
Since the band/gap ratio for the chosen spectral scaling is proportional to $ N e^{-N}$, it vanishes 
as  $ N \to \infty$ and the thermodynamic limit of the stochastic soliton lattices (\ref{turb}) can be associated with one-dimensional soliton gas. 
This also supports our interpretation of the thermodynamic limit of the Whitham equations as the kinetic equation for solitons.

\section{Averaged conservation laws}
Integrating (\ref{tWh}) by $E$ over the interval $(-1,0)$ we obtain the thermodynamic limit of the wave conservation law:
\be \label{}
\partial _T \kappa = \partial_X \Omega \, .
\ee
where $\pi \kappa$ is the full density of waves (integrated wavenumber) (\ref{kappa}) and $\pi \Omega$ is the corresponding
'integrated  frequency'(see (\ref{Omega})).

The quasimomentum $p_N(E)$ is known to be the generating function for the Kruskal integrals $I_k^{(N)}$
\cite{DN}, \cite{JM}:
\be \label{}
p_N(E)=2\sqrt{-E}(1+\sum \limits^{\infty}_{k=0}\frac{I_k^{(N)}}{(-2E)^k}) \, , \qquad -E \gg 1 \, ,
\ee
\be \label{kr}
I_0^{(N)}=\lim \limits_{L \to \infty} \frac{1}{L}\int \limits ^L_0 u_N(x)dx \, ,  \ \ 
I_1^{(N)}=\lim \limits_{L \to \infty} \frac{1}{L}\int \limits ^L_0 \frac{1}{2}u_N^2(x)dx \, , \dots
\ee
Then the thermodynamic limit for the Kruskal integrals is calculated as
$$
 P_k(X,T)= T\hbox{-}\lim I_k^{(N)}= T\hbox{-}\lim\left( \hbox{Res}_{ -\infty}[ \frac{2^{2k+1}}{2k+1}\frac{dp_N}{dE}
(-E)^{\frac{2k+1}{2}}]\right)  
$$
$$
=\frac{2^{2k+1}}{(2k+1)\pi}(-1)^{k+1}\int \limits^{0}_{-1}(-E)^{\frac{2k+1}{2}}\alpha'(E;X,T)dE $$
\be \label{tkrus}
=\frac{2^{2k+1}}{2k+1}(-1)^{k+1}\int \limits^{1}_{0}\eta^{2k+1}f(\eta; X,T)d\eta \, , \qquad k=0,1,2, \dots
\ee
Thus, the infinite set of the {\it independent} averaged conservation laws takes the form
\be \label{cons}
\partial_T P_k=\partial_X Q_k \, , \qquad k=0,1,2 \dots  ,
\ee
where the 'fluxes' $Q_k(X,T)$ are found analogously to (\ref{tkrus}) from the thermodynamic limit for the quasienergy
differential (\ref{tlim}):
\be \label{tflux}
Q_k(X,T)= \frac{2^{2k+1}}{2k+1}(-1)^{k+1}\int \limits^{1}_{0}\eta^{2k+1}f(\eta;X,T)s(\eta; X,T)d\eta \, .
 \ee
In particular, it follows from (\ref{kr}), (\ref{tkrus}) that the thermodynamic limits for the first two KdV moments are
\be \label{mom}
\overline u (X,T)= -2 \int \limits_{0}^{1}\eta f(\eta; X,T)d\eta \, ,  \qquad
\overline {u^2}(X,T) = \frac{16}{3} \int \limits^{1}_{0}\eta^3 f(\eta; X,T)d\eta \,  .
\ee

We note in conclusion, that it is clear that the procedure of the thermodynamic limiting transition can be extended to other integrable Whitham hierarchies. 
Another implication of the developed theory is the thermodynamic limit for the Krichever algebro-geometrical procedure of integration 
of the Whitham equations \cite{Kr}, \cite{DN}.  This result  will be published in a separate paper \cite{E}.

\vspace{0.5cm}
{\bf Aknowledgements}

I would like to thank Alexander Krylov for his permanent interest in this work and numerous useful discussions.
I am also grateful to Martin Kruskal for his inspiring comments at the NEEDS 2002 meeting where part of the
results of this work has been reported.


\begin{thebibliography}{40}

\bibitem{Wh} G.B. Whitham,  Proc.Roy.Soc. {\bf A283} (1965) 238.

\bibitem{FFM}  H. Flaschka, G. Forest, D.W. McLaughlin, Comm. Pure Appl.
Math. {\bf 33} (1979) 739.

\bibitem{LL} P.D.Lax and C.D.Levermore, Comm. Pure Appl. Math. {\bf 36}
(1983) 253,571,809.

\bibitem{V87} S.Venakides, T. Am.Math.Soc. {\bf 301}(1987)189

\bibitem{V90} S.Venakides,  Comm. Pure Appl. Math. {\bf 43}
(1990) 335.

\bibitem{DVZ} P. Deift , S. Venakides, and X. Zhou,  International Math. Research.
Journ., no 4, (1997) 285.

\bibitem{V89} S. Venakides, Comm.Pure Appl.Math. {\bf 42} (1989) 711.

\bibitem{JM} R.Johnson and J.Moser, Comm.Math.Phys. {\bf 84} (1982) 403.

\bibitem{EKMV}G.A.El, A.L.Krylov, S.Molchanov, S.Venakides, Physica D {\bf 152-153}
653

\bibitem{Zakh} V.E.Zakharov, Sov.Phys.JETP {\bf 60} (1971) 1012.

\bibitem{DN} B.A.Dubrovin and S.P.Novikov, Russian  Math. Surveys {\bf 44}
(1989) 35 .

\bibitem{N} S.P. Novikov, Func. Anal.Pril. {\bf 8} (1974) 54.

\bibitem{L} P.D. Lax, Comm.Pure Appl. Math. {\bf 26} (1975) 141.

\bibitem{IM} A.R.Its and V.B.Matveev, Theor.Math.Phys. 23 (1975) 343.

\bibitem{WK} M.I.Weinstein, J.B.Keller, SIAM Jornal Appl.Math. {\bf 47}(1987)941.

\bibitem{GMZ} A.V. Gurevich , N.G. Mazur, K.P. Zybin JETP {\bf 90} (2000) 695.

\bibitem{KE} A.L.Krylov, G.A. El, Soviet Math. Surveys {\bf 54} (1999) 439.

\bibitem{Kr} I.M.Krichever , Func. Anal. Appl. {\bf 22} (1988) 200.

\bibitem{E} G.A.El, Algebro-geometrical solutions of the infinite-genus Whitham eqations,
(to be published).

\end{thebibliography}
\end{document}